# 30th Anniversary of the EMC Effect

*Thirty years on, CERN's EMC effect still puzzles experimentalists and theorists.*

*Douglas Higinbotham, Jefferson Laboratory; Gerald A. Miller, University of Washington; Or Hen, Tel Aviv University and Klaus Rith, University of Erlangen-Nürnberg*

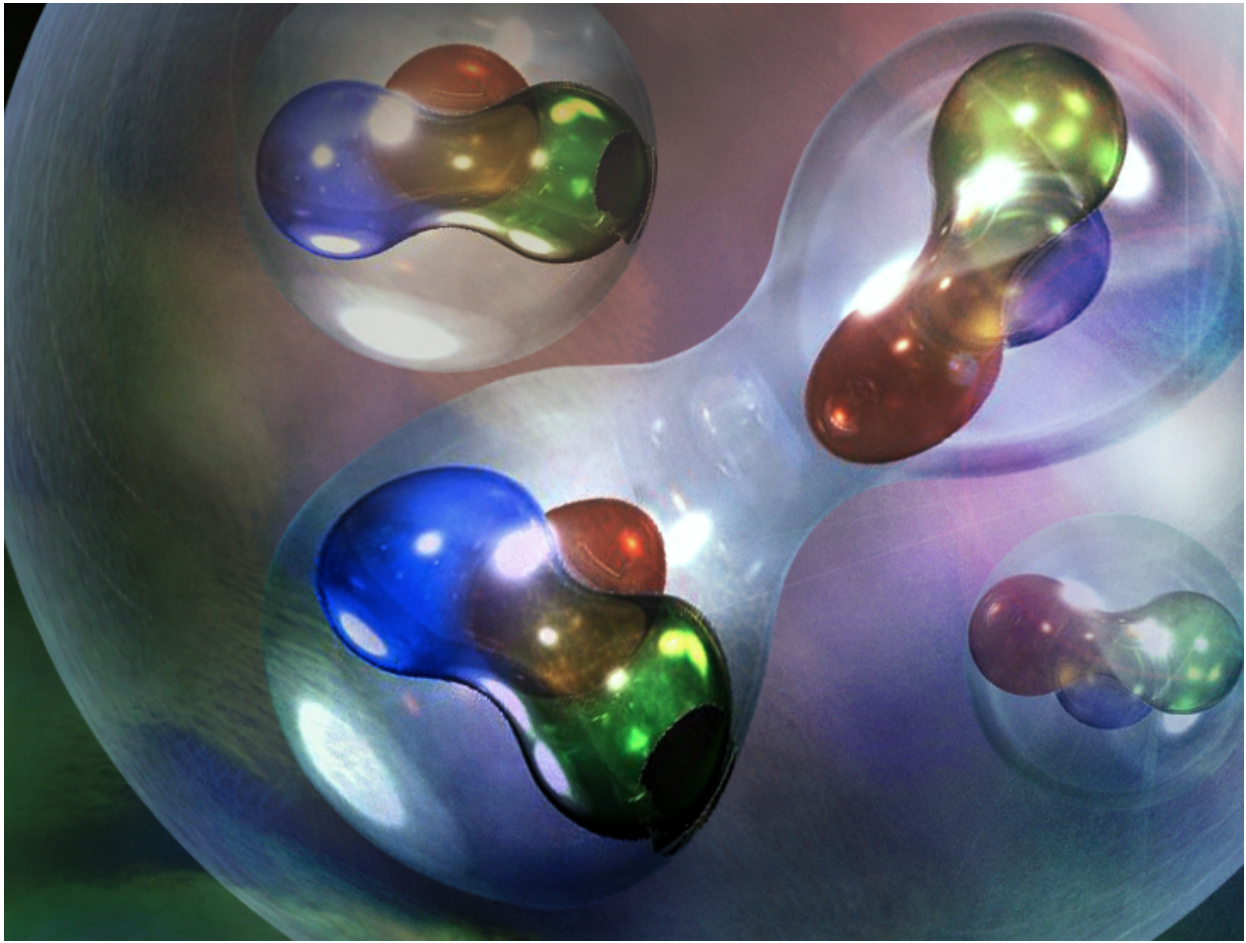

# 30th Anniversary of the EMC Effect

*Article Summary Statement:*
*Thirty years on, CERN's EMC effect still puzzles experimentalists and theorists.*

Contrary to stereotypes, advances in science aren't typically about yelling, "Eureka!"; they are about the results that make the researcher say, "That's strange." This is, in fact, what happened thirty years ago when the CERN European Muon Collaboration (EMC) took the ratio of their per-nucleon deep-inelastic muon scattering data on iron to deuterium.

These data were plotted as a function of Bjorken x, which in deep-inelastic scattering is interpreted as the fraction of the nucleon's momentum carried by the struck quark. Since the binding energies of nucleons in the nucleus are several orders of magnitude smaller than the momentum transfers of deep-inelastic scattering, naively, one would expect such a ratio to give a result of unity, save for small corrections for the Fermi motion of nucleons in the nucleus. What the EMC experiment revealed was an unexpected downward slope to the ratio as printed in the CERN Courier in November of 1982, as shown in Fig. 1, and then in a refereed journal in March 1983 (J.J. Aubert et al.,1983).

This surprising result was confirmed by many groups, culminating with the high-precision electron and muon scattering data from ( SLAC (Gomez et al., 1994), Fermilab (Adams et al., 1995), and NMC at CERN (Amaudruz et al., 1995 and Arneodo et al., 1996) with representative data shown in Fig. 2. The conclusion from the combined experimental evidence was that the effect had a universal shape, was independent of the squared four-momentum transfer $Q^2$, increased with nuclear mass number A, and scaled with the average nuclear density.

The primary theory interpretation of the EMC effect, the x>0.3 region, was simple: quarks in nuclei move throughout a larger confinement volume, and as the uncertainty principle implies, carry less momentum than quarks in free nucleons. The reduction of the ratio at even lower x, named the shadowing region, was attributed either to the hadronic structure of the photon or, equivalently, to the overlap in longitudinal direction of small-x partons from different nuclei. These notions gave rise to a host of models: bound nucleons are larger than free ones; quarks in nuclei move in 6, 9, and even up to 3A quark bags. More conventional explanations, such as the influence of nuclear binding, enhancement of pion cloud effects, and a nuclear pionic field, were successful in reproducing some of the nuclear deep-inelastic scattering data.

One could even combine different models to produce new ones; this led to a plethora of models that reproduced the data (Geesaman et al., 1995), causing one of the authors of this article to write that EMC means Everyone's Model is Cool. It is interesting to note that none of the earliest models were very much concerned with the role of two-nucleon correlations, except as relating to six-quark bags.

The initial excitement was tempered as deep-inelastic scattering became better understood and data became more precise. Some of the more extreme models were ruled out by their failure to match well-known nuclear phenomenology. Moreover, inconsistency with the baryon momentum sum rules led to the downfall of many others. Since some models predicted an enhanced nuclear sea, the nuclear Drell-Yan process was suggested as a way to disentangle the various possible models. In this process, a quark from a proton projectile annihilates with a nuclear anti-quark to form a virtual photon, which, in turn, becomes a leptonic pair (R. Bickerstaff et al., 1984). The experiment was done and none of the existing models provided an accurate description of both sets of data, a challenge that remains to this day (D. Alde et al., 1984.)

A significant shift in our experimental understanding of the EMC effect occurred when data on $^9$Be became available (Seely et al., 2009). These new data changed the experimental conclusion that the EMC effect data follows the average nuclear density, and instead suggested that the effect follows local nuclear density. In other words, even in deep-inelastic kinematics, $^9$Be seems to act like two alpha particles with a single nearly free neutron, rather than like a collection of nucleons whose properties are all modified.

This led experimentalists to ask if the $x_B > 1$ scaling plateaus (CERN Courier http://cerncourier.com/cws/article/cern/29472 ) that have been attributed to short-range nucleon-nucleon correlations, a phenomenon also associated with high local densities (CERN Courier http://cerncourier.com/cws/article/cern/37330 ), could be related to the EMC effect. Figure 3 shows the kinematic range of the EMC effect along with the $x > 1$ short-range correlation (SRC) region. While the dip at *x*=1 has been shown to vary rapidly with $Q^2$, the EMC effect and the magnitude of the *x*>1 plateaus are basically constant within the $Q^2$ range of the experimental data. Plotting the slope of the EMC effect, 0.3<*x*<0.7, against the magnitude of scaling *x*>1 plateaus for all the available data, as shown in Fig. 4, revealed a striking correlation (Weinstein et al., 2011). This phenomenological relationship has led to renewed interest in understanding how strongly correlated nucleons in the nucleus may be affecting the deep-inelastic results.

In February 2013, on nearly the thirtieth anniversary of the EMC publication, experimentalists and theorists came together at a special University of Washington Institute of Nuclear Theory workshop ( http://www.int.washington.edu/PROGRAMS/13-52w/ ) to review our understanding, discuss recent advances, and plan new experimental and theoretical efforts. In particular, an entire series of EMC and SRC experiments are planned for the new 12 GeV electron beam at Jefferson Lab and analysis is underway of new Drell-Yan experimental data from Fermilab.

Although the EMC effect is now 30 years old, the new experimental results have given new life to this old puzzle, and no longer is Every Model Cool. Understanding the EMC effect implies understanding how partons behave in the nuclear medium, and thus has far-reaching consequences not only for the extraction of neutron information from nuclear targets, but also for understanding such effects as the NuTeV anomaly (CERN Courier http://cerncourier.com/cws/article/cern/40108 ) or the observed excesses in the neutrino cross sections at MiniBooNe (CERN Courier http://cerncourier.com/cws/article/cern/29877 ).

**Further Reading**

**About the authors**

Douglas Higinbotham, Jefferson Laboratory; Gerald A. Miller, University of Washington; Or Hen, Tel Aviv University and Klaus Rith, University of Erlangen-Nürnberg

**Figure Captions**

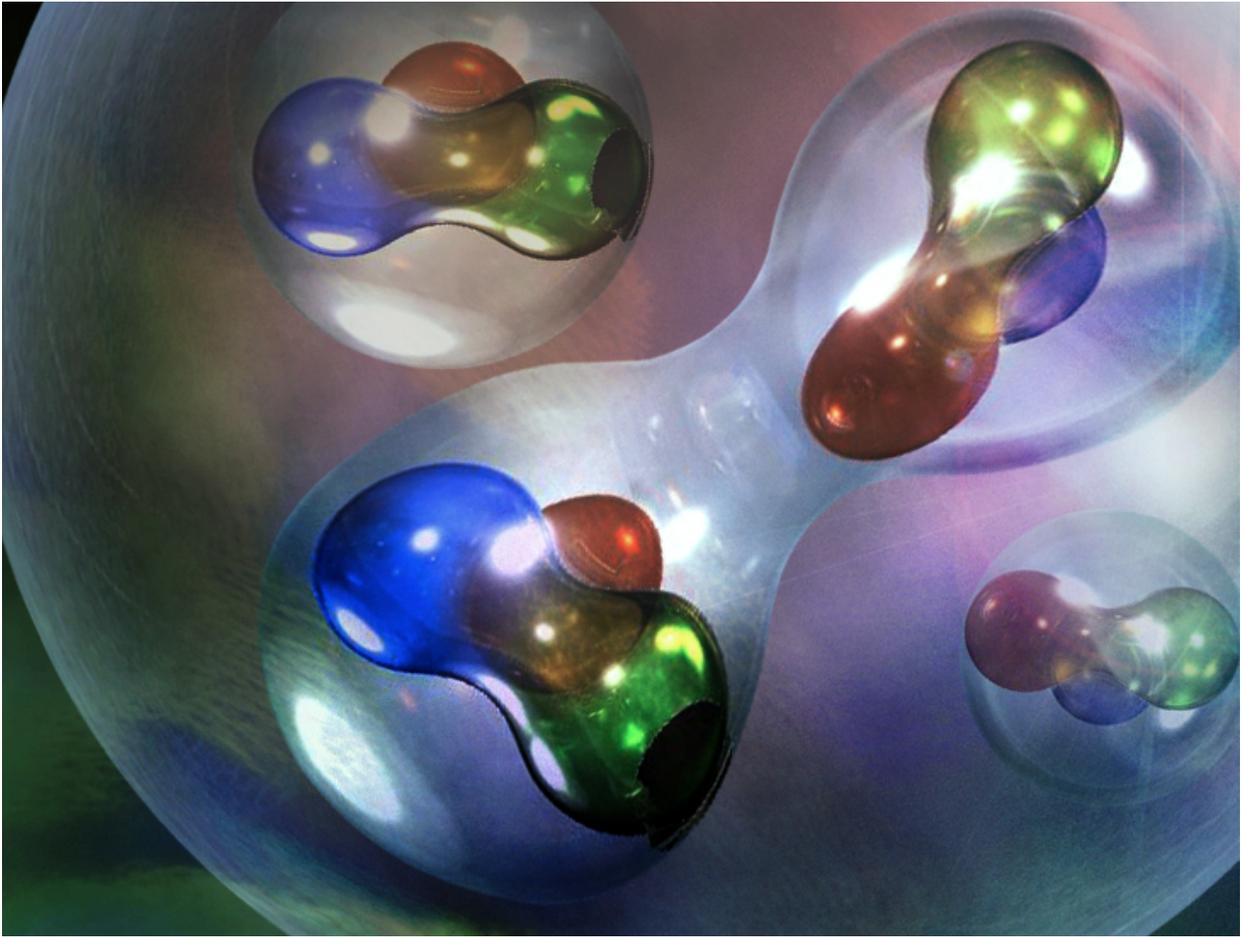

"Cover-Art":   An artist's depiction of nucleons being distorted in the nuclear medium as they come close together.

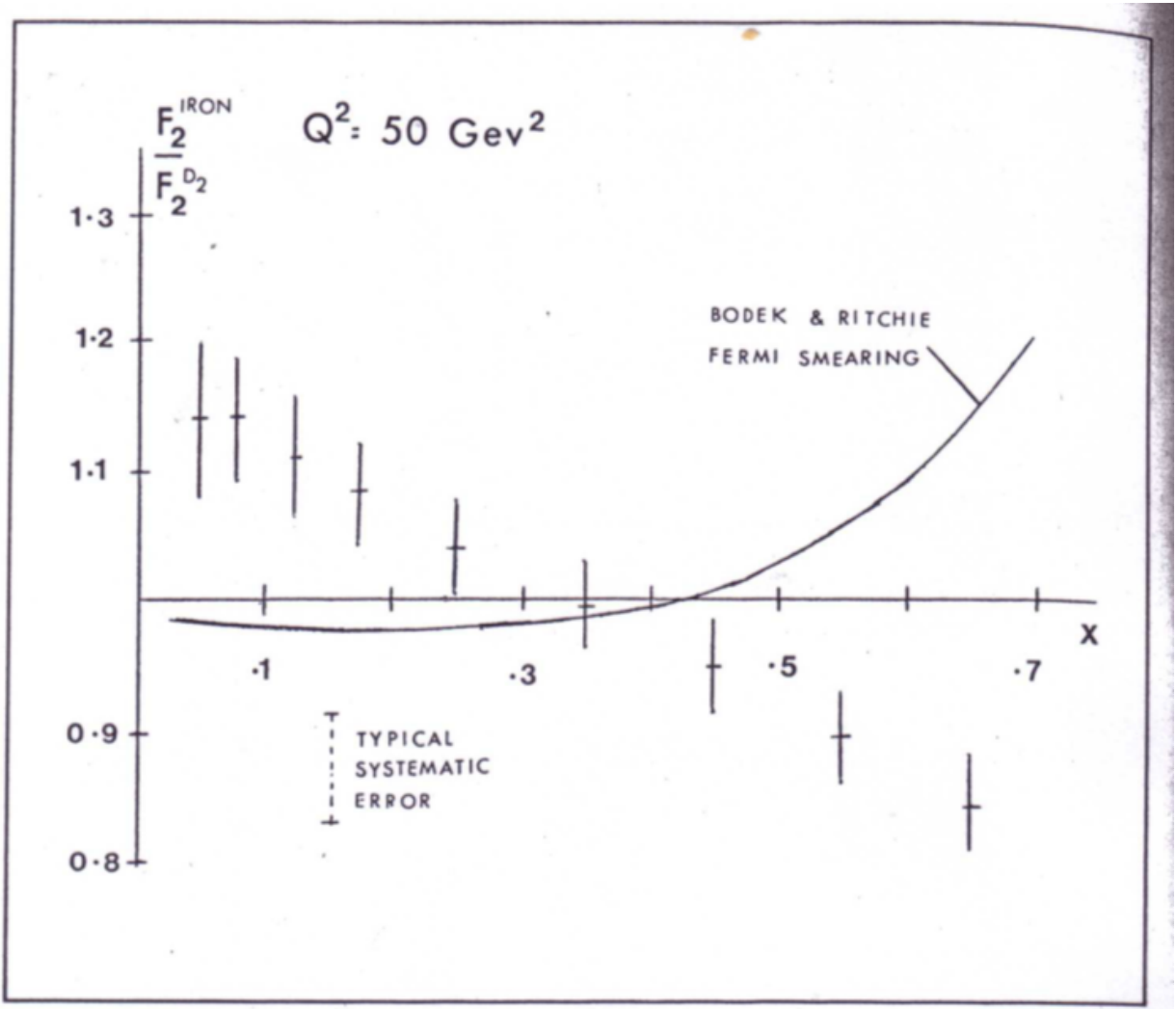

Fig. 1: Image of the EMC data as it appeared in the November 1982 issue of the CERN Courier. This image nearly derailed the highly cited refereed publication (Aubert et al., 1983), as the editor argued that the data had already been published.

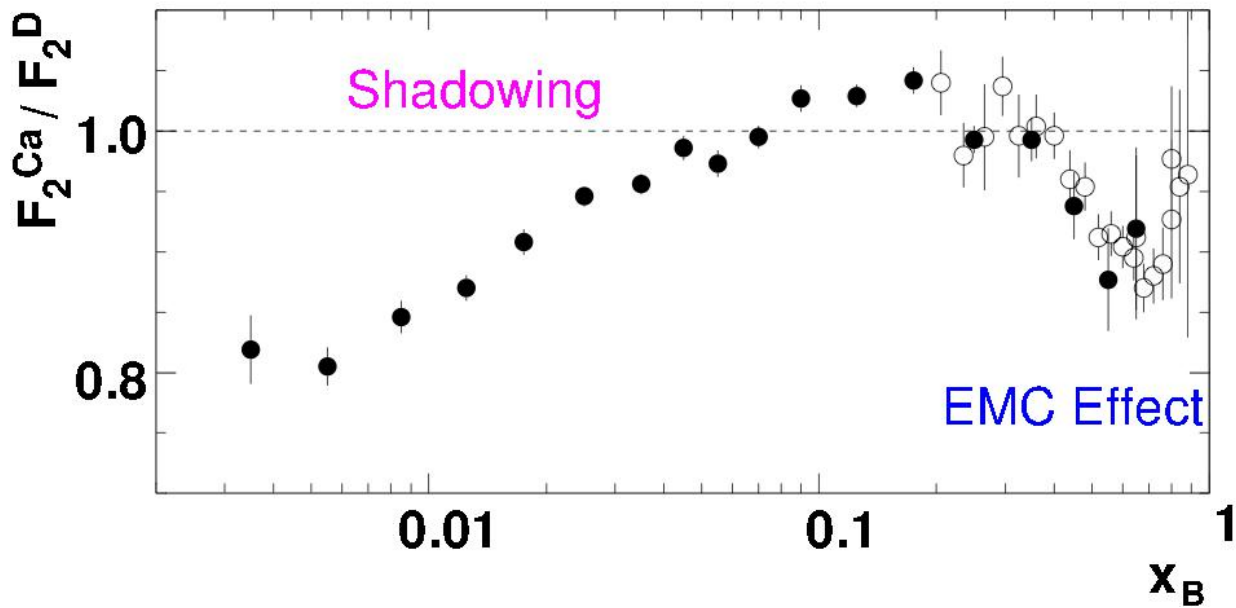

Fig. 2: The image shows the ratio of deep-inelastic cross sections of Ca to D from NMC (solid circles) and SLAC (open circles). The downward slope from 0.3 < x < 0.7 and subsequent rise from $x_B > 0.7$ is a universal characteristic of EMC data and has became known as the EMC effect. The reduction of the ratio at lower values of $x_B$, where valence quarks should no longer be playing a significant role, is known as the shadowing region.

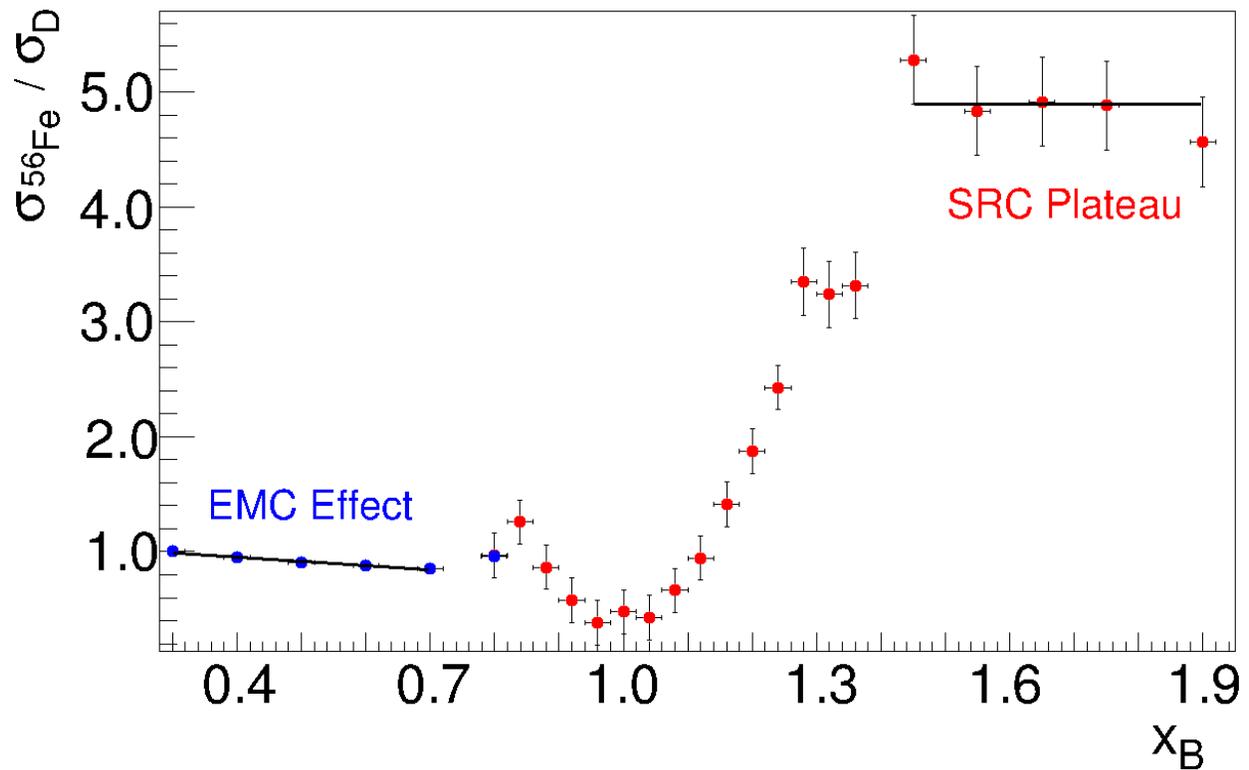

Fig. 3: While the previous plot focused on lower values of $x_B$, this one focuses on the valence quark region. In this region, the EMC effect slope from $0.3 < x < 0.7$ and the $x > 1$ plateaus due to nucleon-nucleon SRC can clearly be seen. Both the EMC effect slope and the SRC plateaus are rather $Q^2$ independent, while the dip at $x = 1$ fills in as $Q^2$ increases.

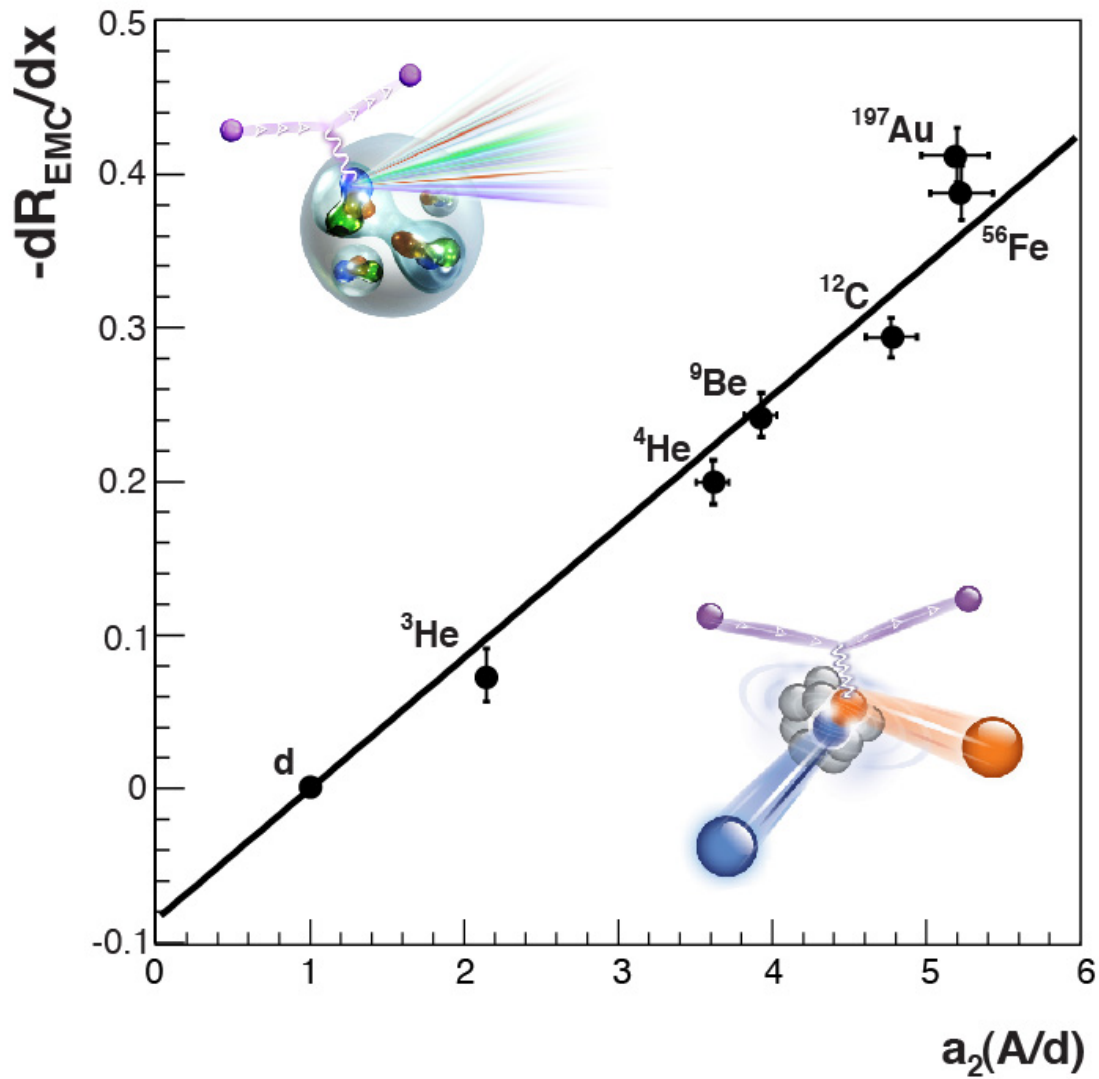

Fig. 4: The slope of the EMC effect, dR/dx for $0.3 < x < 0.7$ with $R = F^2_A/F^2_D$, is plotted versus the magnitude of the observed $x > 1$ plateaus, denoted as $a_2$, for various nuclei. For data that were taken by completely different groups, the linearity is striking and has caused renewed interest in understanding the cause of both effects. The inset cartoons illustrate the kinematic difference of deep inelastic EMC effect scatterings and the scattering from a correlated pair in $x > 1$ kinematics.